\documentclass[12pt,a4paper]{article}

\usepackage{times}
\usepackage{graphicx}
\usepackage{euscript,amssymb}
\usepackage{amsfonts}
\usepackage{amssymb,amsmath}
\usepackage{mathrsfs}
\usepackage{fancybox}
\usepackage[latin1]{inputenc}
\usepackage{pst-all}
\usepackage{epsfig}
\usepackage{rotating}
\usepackage{subfig}
 
\newcommand{\be}{\begin{equation}}
\newcommand{\ee}{\end{equation}}
\newcommand{\bea}{\begin{eqnarray}}
\newcommand{\eea}{\end{eqnarray}}
\newcommand{\lb}{\label}
\newcommand{\bdm}{\begin{displaymath}}
\newcommand{\edm}{\end{displaymath}}
\newcommand{\D}{{\rm d}}
\newcommand{\E}{{\rm e}}

\def\vec#1{\mathbf{#1}}

\begin{document}

\begin{titlepage}

\noindent
\begin{center}
\vspace*{1cm}

{\large\bf ON A QUANTUM WEYL CURVATURE HYPOTHESIS}\footnote{Invited
  contribution for a special topical collection celebrating Sir Roger
  Penrose's Nobel Prize, edited by I.~Fuentes and H.~Ulbricht.}

\vskip 1cm

{\bf Claus Kiefer} 
\vskip 0.4cm
 Faculty of Mathematics and Natural Sciences, University of Cologne,\\
 Institute for Theoretical Physics, Cologne, Germany
\vspace{1cm}

\begin{abstract}
Roger Penrose's Weyl curvature hypothesis states that the Weyl
curvature is small at past singularities, but not at future
singularities. We review the motivations for this conjecture and
present estimates for the entropy of our Universe. We then extend
this hypothesis to the quantum regime by demanding that the initial
state of primordial quantum fluctuations be the adiabatic vacuum in a
(quasi-) de~Sitter space. We finally attempt a justification of this
quantum version from a fundamental theory of quantum gravity and
speculate on its consequences in the case of a classically
recollapsing universe.
\end{abstract}

\end{center}

\end{titlepage}


\section{How special is our Universe?}

Our Universe seems to be very special. On large scales, it is
approximately homogeneous and isotropic; as indicated by the cosmic
microwave background (CMB), initial anisotropies are limited by a
number of order $10^{-5}$. The Universe is also very young. The
observed age of about $13.8$ billion years may not seem small on
everyday standards, but it is surprisingly small when compared to
other scales; Poincar\'e cycles, for example, are much bigger even for very
small systems. 
In fact, $13.8$ billion years is more or less the minimal time
needed for  main sequence stars, habitable planets, and life to
develop.

Can one estimate quantitatively {\em how special} the Universe is?
An answer can be provided by calculating its entropy in an appropriate way
and comparing it with the maximum possible entropy. 
For ordinary matter, states of maximum entropy are homogeneous, so one
might wonder whether the Universe started in a state of high entropy.
That this argument is misleading comes to light when one takes into
account the contribution of {\em gravitational} degrees of freedom to
entropy. Gravity is universally attractive; consequently,
inhomogeneous (`condensed') states are entropically preferred. Unfortunately, no
general expression for the entropy of the gravitational field is
known. But what we know is an exact formula for the entropy of a black hole,
which is arguably the most condensed system in nature. This formula
is the expression for the Bekenstein--Hawking entropy and is given by
\be
\lb{SBH}
S_\mathrm{BH} = k_\mathrm{B} \frac{Ac^3}{4 G \hbar}\equiv
 k_\mathrm{B} \frac{A}{(2l_{\rm P})^2},
\ee
where $A$ denotes the area of the black hole's event horizon, and
$l_{\rm P}=\sqrt{G\hbar/c^3}$ is the Planck length. 
If we set Boltzmann' constant $k_\mathrm{B}$ equal to one (as we shall
do in most of the coming equations), the Bekenstein--Hawking entropy
gives the area in terms of (twice) the Planck length squared.
In the special case of a spherically-symmetric (Schwarzschild) black
hole with mass $M$, Eq.~(\ref{SBH}) assumes the form
\be
\lb{SBH-SS}
S_\mathrm{BH}\approx 1.07 \times 10^{77}k_{\rm B}\left(\frac{M}{M_{\odot}}\right)^2,
\ee
where $M_{\odot}$ is the solar mass. 

Equation (\ref{SBH}) is found using the laws of black-hole mechanics.
Its statistical origin is, so far, unknown, despite many attempts (and
preliminary results) using current
approaches to quantum gravity; see, for example, Kiefer (2012a) and the
references therein. Following John Wheeler's old idea of ``it from
bit'', one can divide the area $A$ into cells of size Planck-mass
squared and calculate the number of ways one can attach the bits $0$
and $1$ to these cells. This gives a simple model to understand the
possible origin of (\ref{SBH}), and it also provides the means to
calculate statistical correction terms of the form
$\propto\ln(A/l_{\rm P}^2)$, which arise from Stirling's
formula for factorials (Kiefer and Kolland~2008). Except for
Planck-size black holes, these
corrections terms are negligible and will not be taken into
account below.

Some time ago, Roger Penrose made use of (\ref{SBH-SS}) to estimate
the maximum possible entropy of our Universe; see Penrose (1977, 1979,
1981, 1986). For this purpose, he assumed that all matter in the
observable Universe were assembled into one gigantic black hole. The
size of the observable Universe is defined by the present size of the
particle horizon. Using
(\ref{SBH-SS}), Penrose obtained a value of the order
$10^{123}$ (Penrose~1981). (From now on, we set $k_{\rm
    B}=1$.) Taking into account the fact that our Universe is
presently accelerating and that we thus have to use a
Schwarzschild--de~Sitter solution (with a value for the cosmological
constant $\Lambda$ inferred from the \textsc{Planck} data) instead of a Schwarzschild solution,
this value is reduced to (Kiefer~2012b) 
\be
\lb{Smaxnew}
S_{\rm max}\approx 1.8\times 10^{121} . 
\ee

But for an accelerating Universe this is, in fact, not the maximum
possible entropy. 
For the observed value of $\Lambda$, there is a contribution from the 
event horizon of the de~Sitter space, which will be the late-time geometry
of our Universe (under the assumption of a constant $\Lambda$ and not
a time-dependent dark energy). The general expression for this entropy was derived
by Gibbons and Hawking (1977) and reads
\be
\lb{SEH}
S_{\rm EH}= \frac{3\pi}{\Lambda l_{\rm P}^2}.
\ee
Taking into account the present uncertainties in the cosmological data, Egan and
Lineweaver (2010) found from this the following numerical value:
\be
\lb{SEH-value}
S_{\rm EH}\approx 2.88\pm 0.16\times 10^{122}.
\ee
Comparing (\ref{SEH-value}) with (\ref{Smaxnew}), it is clear that a
future de~Sitter space is entropically preferred by about one order of
magnitude over a state with all matter being
assembled into one gigantic black hole.

In order to calculate the probability for our Universe, the maximum
value (\ref{SEH-value}) must be compared with the present value for
the entropy within the same region. Egan and Lineweaver (2010) 
present a detailed estimate of all relevant contributions to the
entropy, both for the observable Universe (their Table~1) and for the
matter within the event horizon (their Table~2). The dominating
contributions from the gravitational side are supermassive black holes
(SMBHs), followed by stellar black 
holes.  
For the region inside the event
horizon, the authors present the value
\be
S_{\rm SMBH}=1.2^{+1.1}_{-0.7}\times 10^{103}
\ee
for the entropy from supermassive black holes, while the biggest
contribution to non-gravitational entropy comes from the CMB photons,
with the value
\be
S_{\rm CMB}=2.03\pm 0.15 \times 10^{88},
\ee
and a slightly smaller value for the entropy of relic neutrinos.
We see that the non-gravitational entropy is completely
negligible. (Already the entropy of the black hole in the
  centre of the Milky Way is about hundred times the entropy of the
  CMB photons.)

With these numbers, following Penrose (1981), we can estimate the
probability of our Universe as follows:
\be
\lb{Penrose-estimate}
\frac{\exp(S_{\rm SMBH})}{\exp(S_{\rm max})}\approx \frac{\exp(1.2\times
  10^{103})}{\exp(2.88\times 10^{122})}\approx \exp(-2.88\times
10^{122}).
\ee
In the ratio of these two `multillions' [a term used by Eddington
for double exponentials such as ${10^{10}}^{10}$, see Eddington (1931,
p.~450)], the huge number in the numerator is completely negligible
compared to the even huger number in the denominator. A similar
argument applies to the case with one black hole using (\ref{Smaxnew}).

As we see from (\ref{Penrose-estimate}), our Universe is very special
indeed. From a pure entropic point of view, one would have expected
that the Universe started from a very inhomogeneous state with black
holes or already from a de~Sitter-type space with large event
horizon. A smooth initial state without cosmic event horizon is
extremely special. Since this would correspond to vanishing Weyl tensor,
Penrose came up with the hypothesis that a fundamental theory should
predict vanishing Weyl tensor at past singularities. In 
Penrose (1986, p.~138, italics in the original), he uses the following
words:
\begin{quote}
\textsc{Hypothesis (Classical)}: {\em The Weyl curvature vanishes at all past
  singularities, as the singularity is approached from future
  directions.} [This condition can be weakened by only demanding
   that the Weyl tensor be finite, rather than diverging, see
  Penrose (2011, p.~134).]
\end{quote}
He continues by writing: `This has the advantage that white holes, with their
unpleasant anti-thermodynamic behaviour, are excluded. \ldots This
hypothesis is time-asymmetric, as indeed could have been anticipated,
since it yields the time-asymmetric Second Law.'
The Weyl curvature hypothesis (WCH) thus excludes the
presence of white holes. For various aspects of the WCH, see
  the recent essay by Hu (2021) and the references therein.

Of particular interest are the consequences of the WCH for a recollapsing universe
(as is currently disfavoured by observations, but is still a
theoretical possibility). In Fig.~1 we present a diagram similar
to the one presented in Penrose (1981). The `stalactites' there
symbolize black holes (before evaporation); a more probable universe
would have `stalactites' as well as `stalagmites', the latter
representing white holes. (These must not be confused with
  primordial black holes which would correspond to `very long'
  stalactites almost touching the big bang line.) 

\begin{figure}[t]
\label{fig_berlin05_7}
\begin{center}
  \includegraphics[width=10cm]{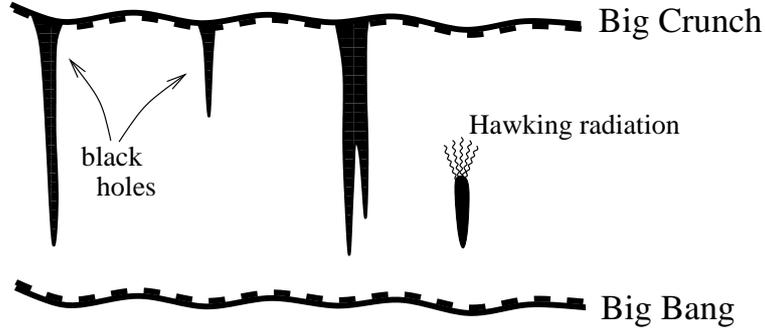}
  \caption{Situation for a recollapsing universe when implementing the
  WCH: the big crunch 
is fundamentally different from the big bang because the big bang is
very smooth (small Weyl tensor, low entropy), whereas the big crunch is very
inhomogeneous (diverging Weyl tensor, high entropy).}
\end{center}
\end{figure}

Vanishing Weyl tensor entails, in particular, the absence of gravitational
radiation. From the WCH it then follows that all gravitational waves must be
retarded. This is analogous to the {\em Sommerfeld condition} stating
the absence of advanced electromagnetic radiation; see Zeh
(2007). Such a condition is crucial for understanding the origin of the
arrow of time. 

Gravitational waves can be described by certain Weyl scalars
constructed from the Weyl tensor (Newman and Penrose~1962). One of
them is
\be
\lb{N-P}
\Psi_4:= -\frac{1}{8c^2}\left(\ddot{h}_+-{\rm i}\ddot{h}_{\times}\right),
\ee
where $h_+$ and $h_{\times}$ denote the two polarization states of weak
gravitational waves. The Newman--Penrose quantity $\Psi_4$ describes
the helicity state $s=-2$, while its complex conjugate describes
$s=+2$. Weyl scalars will play a role in the quantum version of the
Weyl curvature hypothesis below where we will demand that $\Psi_4$ be
small. 

The problem connected with the WCH is thus to understand initial
conditions in cosmology. This was already emphasized by Eddington
(1931). He envisaged the possibility that a low-entropy state is
generated by an extremely improbable fluctuation, which is an idea dating back
to Boltzmann. He called such a process {\em anti-chance}, but was
unwilling to accept this possibility in reactions between atoms or
other physical systems. He saw the only possibility for such a process
in the boundary conditions: ``Accordingly, we 
sweep anti-chance out of the laws of physics---out of the differential
equations. Naturally, therefore, it reappears in the boundary
conditions \ldots'' Among his arguments to reject the idea of an
unlike fluctuation he used a concept that today is known as
`Boltzmann brain': if we just emerged from a gigantic
fluctuation, it would be more likely that `we' would just emerge as
brains seeing a disorganized world rather than an ordered world such as
ours. This is because observing a disorganized world (and even
a partly organized world full of inconsistent documents) is immensely
more probable than observing an organized world.
Eddington speaks of mathematical physicists
instead of Boltzmann brains: ``\ldots it is practically certain that a
universe containing mathematical physicists will at any assigned date
be in the state of maximum disorganisation which is not inconsistent
with the existence of such creatures.'' After rejecting this idea, he
concluded: ``We are thus driven to admit anti-chance; and apparently
the best thing we can do with it is to sweep it up into a heap at the
beginning of time, as I have already described.''

But can we understand this occurrence of anti-chance at the beginning
of time?

\section{Low entropy for early quantum perturbations}

It is not surprising that we know relatively little about the early
phases of our Universe. A generic state would look very different from
our present approximately homogeneous and isotropic world. But given
the fact that symmetries play a fundamental role in physics, one might
speculate that the Universe started with a highly symmetric
state. Alexei Starobinsky came up with the idea that ``the universe
was in a maximum symmetrical state before the beginning of the
classical Friedmann expansion'' (Starobinsky~1979). For this state, he
chose de~Sitter space, which is (as Minkowski space) a state with maximal
symmetry. 

Classical de~Sitter space is homogeneous and isotropic and thus cannot
lead to structure formation. The situation is different if quantum
fluctuations are taken into account. Starobinsky suggested that the
quantum fluctuations for gravitational waves (the gravitons) are
initially in their ground state (the adiabatic vacuum). During the
expansion, vacuum modes with large enough wavelength become excited
and are no longer in their ground state. In the spirit of the
inflationary universe, which was developed in the years after
Starobinsky's suggestion, one can extend this idea also to quantum
scalar modes (scalar components of the metric together with the
inflaton).

In cosmic perturbation theory, one can combine the scalar fluctuations of the
metric and a scalar field into the gauge-invariant
`Mukhanov--Sasaki variable' 
$v(\eta,\vec{x})$, where $\eta$ denotes the conformal time defined by 
${\D\eta}/{\D t} = a^{-1}$, and $a$ is the scale factor of a
Friedmann--Lema\^itre (F-L) universe; see, for example,
  Brizuela {\em et al.} (2016) and the references therein.
We denote the Fourier transform of $v(\eta,\vec{x})$ by $v_{\vec{k}}$; 
we also introduce the Fourier-transformed perturbation variable of
the gauge-invariant tensor perturbations $h_{ij}$ with polarization
$\lambda \in \{+,\times\}$ by
\be
\lb{MS-h}
v_{\vec{k}}^{(\lambda)} := \frac{a\,h^{(\lambda)}_{\vec{k}}}{\sqrt{16\pi G}}.
\ee 
We note that an important feature in the definition of these variables
is the rescaling with respect to $a$. This becomes especially relevant in
quantum theory. By expression (\ref{MS-h}) we can relate the
variable $v_{\vec{k}}^{(\lambda)}$ to the Weyl scalar (\ref{N-P})
(similar features hold for the relation between  $v_{\vec{k}}$ and
other Weyl scalars). 

For a perturbed inflationary universe, one obtains an
action containing a background part with scale factor $a$ and
homogeneous field $\phi$ plus a sum over all $\vec{k}$ with
$\eta$-dependent oscillators described by $v_{\vec{k}}$ and $v_{\vec{k}}^{(\lambda)}$;
these oscillators have the
`frequencies' ${}^\text{S,T}\omega^2_{\vec k}(\eta)$ given by
\be
\label{eq:defomega}
{}^\text{S}\omega^2_{\vec{k}}(\eta)=k^2-
\frac{z^{\prime\prime}}
{z}
\ee
for the scalar (metric and scalar-field) perturbations and by
\be
\label{eq:defTomega}
{}^\text{T}\omega^2_{\vec{k}}(\eta)=k^2-
\frac{a^{\prime\prime}}
{a}
\ee
for the tensor perturbations, respectively; moreover, we have
$z:=\phi'/H$, where $H$ is the Hubble parameter, and primes denote
derivatives with respect to conformal time. (We restrict here
  to minimally coupled fields.)

Since $v_{\vec{k}}$ and $v_{\vec{k}}^{(\lambda)}$ are quantum
variables, they obey Schr\"odinger equations with respect to $\eta$
(or $t$). Assuming now the initial condition that these states be in
their {\em adiabatic vacuum state}, one has for them the wave
functions 
\be
\label{Gaussianansatz2}
\psi_{\vec{k}}(v_{\vec{k}}) =
\mathcal{N}_{\vec{k}}\exp\left(-\frac{1}{2}\,
  \Omega_{\vec{k}}^{(0)}v_{\vec{k}}^2\right), 
\ee
with $\Omega_{\vec{k}}^{(0)}=k$, and a similar expression for the
tensorial modes. With this initial condition, the solution of the
Schr\"odinger equations for the modes are Gaussians of the form
(\ref{Gaussianansatz2}) with an $\eta$-dependent factor
$\Omega_{\vec{k}}^{(0)}(\eta)$. These states are now sums over excited states. In
the case of pure de~Sitter 
inflation the factors in the exponent read
\be \label{om0ds}
{}^{\text{dS}}\Omega^{(0)}_{\vec{k}}(\eta):=\frac{k^3\eta^2}{1+k^2\eta^2}+\frac{{\rm
    i}}{\eta(1+k^2\eta^2)};
\ee
the expression for slow-roll inflation is more complicated (Bizuela
{\em et al.}~2016). The important feature is the occurrence of an
imaginary term in this expression. Quantum mechanically, the ensuing
states correspond to two-mode squeezed states.

From the solutions of the Schr\"odinger equations one can derive the
power spectra for the density perturbations and for the primordial
gravitational waves. The corresponding expressions contain explicitly
the Planck length and could thus be interpreted as
quantum-gravitational effects (Krauss and Wilczek~2014). 

Since the above wave functions are pure states, they have vanishing
entropy (no missing information). A positive entropy comes into play
when considering interactions of the modes with other degrees of
freedom (such as fields from an effective quantum field theory) or
with higher-order perturbations. This is connected with the 
process of decoherence -- the emergence of classical behaviour (Joos
{\em et al.}~2003). These interactions only play a role for the
excited states, not the ground states.
Detailed calculations of decoherence and
entanglement for the primordial fluctuations can be found in Kiefer
{\em et al.} (2007). There the von Neumann entropy
\be
\lb{entropy}
S=-{\rm Tr}(\rho\ln\rho)
\ee
was calculated. Alternatively, one can consider the `linear entropy'
$S_{\rm lin}={\rm Tr}(\rho-\rho^2)$,
which is bounded between zero (pure state) and one (maximally
mixed state) and can be used to quantify the degree of purity in a
simpler way than \eqref{entropy}.

The reduced density matrix $\rho$ occurring in (\ref{entropy}) is
obtained by integrating out irrelevant degrees of freedom from a
totally entangled quantum state containing $a$, $\phi$, the
$v_{\vec{k}}$ (resp. the tensorial modes), and the irrelevant degrees
of freedom. To obtain (\ref{entropy}), one has to perform the trace
over $v_{\vec{k}}$.

The resulting entropy increases with increase in the scale factor
$a$. It does not yet lead to the large entropies presented in the
first section. For this, further entropy-producing processes play a
role (e.g. reheating after inflation). But the important point is that
starting with a low-entropy initial state, one has enough
entropy-generating
capacity to generate an arrow of time. Instead of the 
Weyl curvature hypothesis presented in the first section, one can thus
present here the following quantum version:
\begin{quote}
\textsc{Hypothesis (Quantum)}: {\em The quantum states for the Weyl scalars
  describing scalar and tensor modes assume the form of adiabatic
  vacuum states in a (quasi-) de~Sitter space, as the region of small-enough
  scale factors is approached from future 
  directions.}
\end{quote}
We have here replaced `past singularities' with `region of
small-enough scale factors' because it is generally assumed that
singularities are absent in a quantum theory of gravity.

As in the classical case, this is a conjecture only. Can it be
justified at a more fundamental level?

\section{Justification from quantum gravity?}

So far, we have discussed quantum states for primordial fluctuations
in the background of a Friedmann--Lema\^{\i}tre universe. Since these
include metric perturbations, quantum effects of gravity are already
included. We would,
however, expect that a truly fundamental explanation for the Quantum
Weyl Hypothesis comes from an underlying (not yet known) full quantum
theory of gravity, where no background exists. A very
conservative approach and one especially 
suited for discussing conceptual issues is quantum geometrodynamics,
with the Wheeler--DeWitt equation as its central equation; see, for
example, Kiefer (2012a) for a detailed introduction.

For the case of a quantized F-L universe plus the above discussed
primordial fluctuations described by the gauge-invariant variables
$v_{\vec{k}}$, the Wheeler--DeWitt equation reads 
\bea
\lb{WdW2}
\frac{1}{2}&\Biggl\{&a_0^{-2}\E^{-2\alpha}\left[\frac{1}{m_\text{P}^2}\,
\frac{\partial^2}{\partial\alpha^2}-
\frac{\partial^2}{\partial\phi^2}+2a_0^6\E^{6\alpha}\,\mathcal{V}(\phi)\right]
\\ 
&+& \sum_{\vec{k};\text{S,T}}\left[-\,\frac{\partial^2}{\partial
    v_{\vec k}^2}+
{}^\text{S,T}\omega^2_{\vec k}(\eta)\,v_{\vec{k}}^2\right]\Biggr\}\Psi(\alpha,\phi,\{v_{\vec{k}}\}) = 0; \nonumber
\eea
 see, for example,
Brizuela {\em et al.} (2016) and the references therein. 
Here, $\alpha:=\ln(a/a_0)$, where $a_0$ is a reference scale, and the
`frequencies' ${}^\text{S,T}\omega^2_{\vec k}(\eta)$ are given by 
(\ref{eq:defomega}) and (\ref{eq:defTomega}). 
We choose units where $\hbar=c=1$ and where the Planck mass reads
\be
m_{\rm P}^2 := \frac{3\pi}{2G}.
\ee
We emphasize that the potential terms in (\ref{WdW2}) are
{\em asymmetric} with respect to $\alpha\to -\alpha$. In contrast to almost all
the other fundamental equations in physics, the Wheeler--DeWitt
equation thereby distinguishes a 
direction in (intrinsic) time $\alpha$. 
Inspecting the frequencies \eqref{eq:defomega} and
\eqref{eq:defTomega}, one recognizes that they do not depend on $a$
and $\phi$ for large $k$, that is, for small-wavelength modes. This
is also true in the limit of small $a$. Since the $(a,\phi)$-part
(`minisuperspace part') 
of \eqref{WdW2} then decouples from the perturbation part, one can
naturally impose the following initial condition on the total 
quantum state, with $v_{\vec k}$ (and their tensorial partners)
being in the adiabatic vacuum state (\ref{Gaussianansatz2}), compare
Zeh (2007),
\begin{equation}
\lb{psiBB}
\Psi \quad \stackrel{\alpha \, \to \, -\infty}{\longrightarrow}\
\psi_0(\alpha,\phi)\prod_{\vec{k}} \psi_{\vec{k}}(v_{\vec{k}}) .
\end{equation}
This is a product state, which means that tracing out some of the
degrees of freedom will remain ineffective, that is, it will not lead to
a mixed state; thus, the 
entropy for the $(a,\phi)$-variables remains zero after coarse-graining. 
While the state in the Wheeler--DeWitt equation \eqref{WdW2} is timeless, a
semiclassical or `WKB' time comes into play after a Born--Oppenheimer
type of approximation is being employed; see, for example, the
detailed discussion in Kiefer (2012a). In this limit, the
Schr\"odinger equations for the modes of the last section arise as
approximate equations with respect to the WKB times $\eta$ or $t$. 
For bigger values of $a$, entanglement will emerge, and the state
\eqref{psiBB} is replaced by 
\be
\lb{entangled}
\Psi\big(\alpha,\phi,\{v_{\vec{k}}\}\big) =
\psi_0(\alpha,\phi)\prod_{\vec{k}} \psi_{\vec{k}}(v_{\vec{k}},\eta),
\ee
where the conformal time $\eta$ is to be understood as a function of
$\alpha$ and $\phi$. Here, $\psi_{\vec{k}}(v_{\vec{k}},\eta)$ are
the squeezed states of the last section, which are states of Gaussian
form with the parameter in the exponent given by (\ref{om0ds}) or its
slow-roll generalization.  

There is thus an increase in entanglement entropy from small to
large scale factor and thus from small to large semiclassical time
$\eta$ (or $t$). Within each semiclassical and decohered branch of the
full quantum state, one can express entanglement in terms of
thermodynamic entropy; see, for example, Peres (1995, Chap.~9). The
increase in entanglement entropy could thus be seen as providing the
arrow of time in our Universe.

\begin{figure}[t]
\label{fig_berlin05_8}
\begin{center}
  \includegraphics[width=10cm]{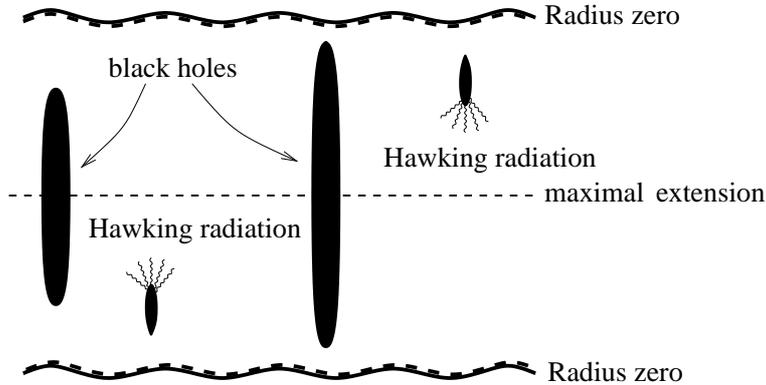}
  \caption{The quantum situation for a ``recollapsing universe'': big crunch
and big bang correspond to the same region in configuration space. The
Weyl tensor is small at both ends.}
\end{center}
\end{figure}

An interesting consequence of this arises for the case of a
classically recollapsing universe (Kiefer and Zeh~1995).
Instead of the classical picture shown in Fig.~1 one arrives at the
quantum picture sketched in Fig.~2. Since the quantum theory does not
distinguish between the regions with a classical big bang and a
classical big crunch (they both correspond to the same region in
configuration space with small $a$), imposing low entropy for the `big
bang' directly leads to low entropy for the `big crunch'. Imposing the
quantum version of the Weyl curvature hypothesis for the region that
would classically be a big-bang singularity would then automatically
entail the same version for the big-crunch region.
Consequently,
the arrow of time would formally reverse near the classical turning
point. But since semiclassical components of the universal wave
function would destructively interfere there, classical systems
are not expected to survive it. Every observer in this quantum universe would thus
only be able to see an expanding universe (Kiefer and Zeh~1995).

This has also consequences for black holes. A time-reversed black hole
is a white hole. Thus, from the point of view of the symmetric picture
shown in Fig.~2, a black hole turns into a white hole after the turning
point. But for real observers, who are subject to the arrow of time
and experience an expanding universe,
there are only black holes.  

That the arrow of time may point in the direction of an expanding
universe, was envisaged long ago. John Wheeler, for example, wrote
(Wheeler~1962, p.~72):

\begin{quote}
The universe is not a system with respect to which ordinary
statistical considerations apply. There is no better evidence on this
point than the correlation between (a) the direction of time in which
entropy increases and (b) the direction of time in which the expansion
of the universe is proceeding.
\end{quote}

These considerations are, of course, speculative. But they are
concrete in the sense that they arise naturally from a straightforward
combination of general relativity with quantum theory, together with a
particular boundary condition. One could investigate similar
conceptual issues in other theories of gravity, for example when a
term proportional to Weyl-tensor squared is added to the
Einstein--Hilbert action; see, for example, 
Kiefer and Nikoli\'c (2017) and the references therein. We leave this
for future work.

\vskip 1cm

\noindent {\em Data Availability Statement}: Data sharing not
applicable--no new data generated.

\vskip 5mm

\noindent The author has no conflicts to disclose.
  


\end{document}